\documentclass[10pt, twocolumn, pre, aps, superscriptaddress, showpacs]{revtex4-1}
\usepackage{amsmath, graphicx, subfigure, tikz}
\begin{document}

\title{Fractal Measures of Sea, Lake, Strait, and Dam-Reserve Shores:\\
Calculation, Differentiation, and Interpretation}
\author{Dilara Y{\i}lmazer}
\affiliation{TEBIP High Performers Program, Board of Higher Education of Turkey, Istanbul University, Fatih, Istanbul 34452, Turkey}
\author{A. Nihat Berker}
\affiliation{TEBIP High Performers Program, Board of Higher Education of Turkey, Istanbul University, Fatih, Istanbul 34452, Turkey}
\affiliation{Faculty of Engineering and Natural Sciences, Kadir Has University, Cibali, Istanbul 34083, Turkey}
\affiliation{Department of Physics, Massachusetts Institute of Technology, Cambridge, Massachusetts 02139, USA}
\author{Y\"ucel Y{\i}lmaz}
\affiliation{Department of Geological Engineering, Istanbul Technical University, Maslak, Istanbul 34469, Turkey}
\affiliation{Faculty of Engineering and Natural Sciences, Kadir Has University, Cibali, Istanbul 34083, Turkey}

\begin{abstract}

The fractal dimensions $d_f$ of the shore lines of the Mediterranean, the Aegean, the Black Sea, the Bosphorus Straits (on both the Asian and European sides), the Van Lake, and the lake formed by the Atat\"urk Dam have been calculated.  Important distinctions have been found and explained.
\end{abstract}
\maketitle

\begin{figure*}[ht!]
\centering
\includegraphics[scale=0.35]{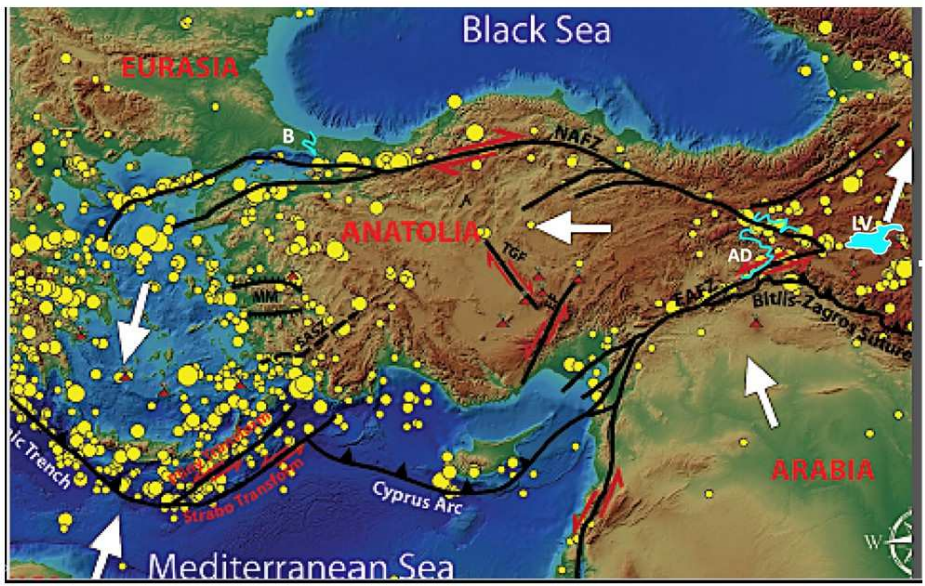}
\caption{Tectonic map of Anatolia and the surrounding regions showing the major faults (black lines) and the earthquake epicenters (yellow circles). The white arrows indicate relative motions of the different regions of the Arabian, Anatolian, and Aegean Plates. The abbreviations are LV for Lake Van, AD for the lake of the Atat\"urk High Dam, and B for the Bosphorus. (Modified after \cite{Yilmaz3}).}
\end{figure*}

\section{Introduction}

\begin{figure}[ht!]
\centering
\includegraphics[scale=0.5]{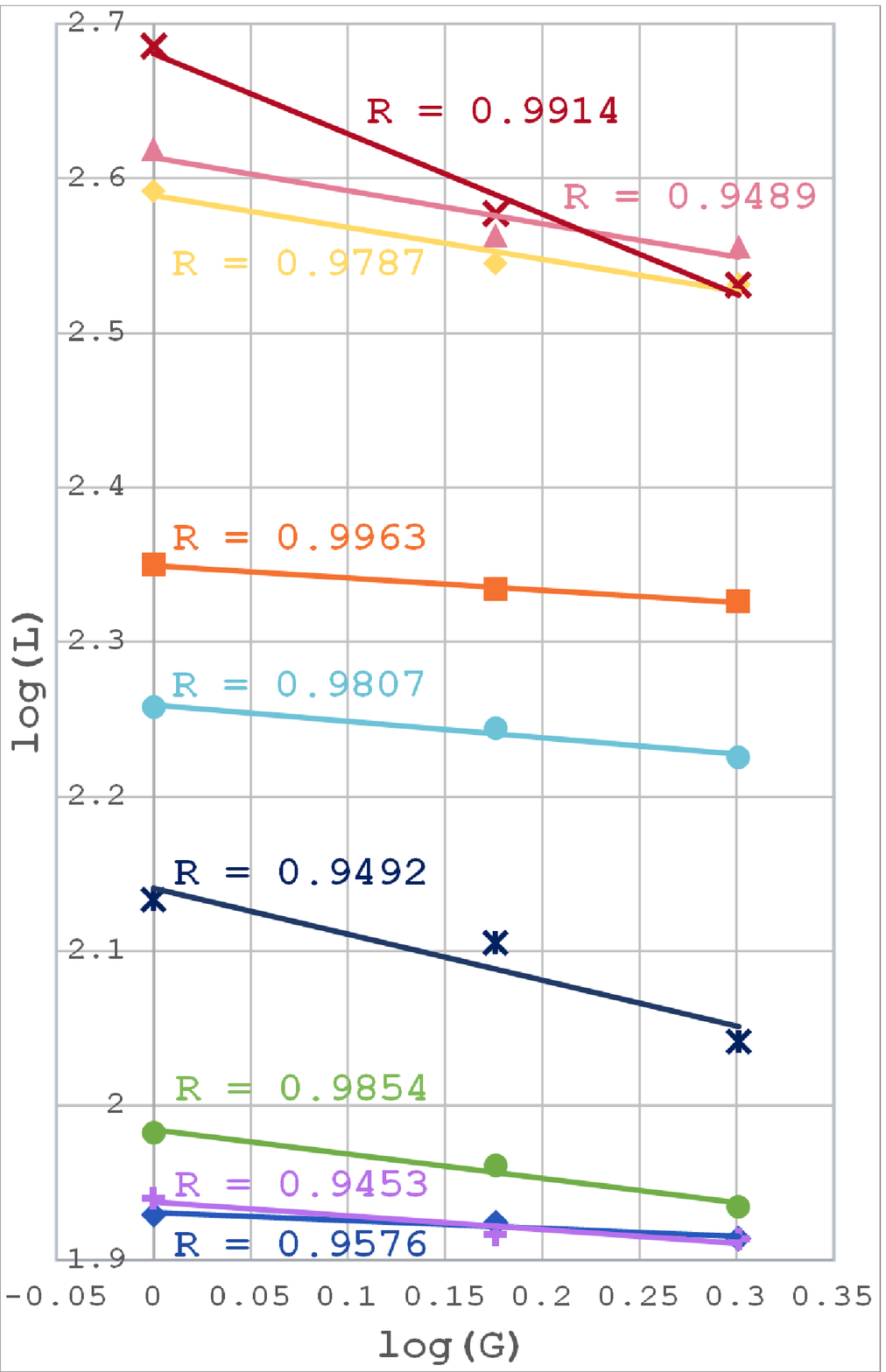}
\caption{Logarithmic plot of the shore length $L$ versus the unit ruler $G$. Each triplet of points this figure represents a large number of measurements, e.g., 840 measurements for the Meditteranean and 907 measurements for the Atat\"urk Dam Lake, as reported in Table I.  The goodness $R$ (see Table I) of the linear fit, shown here, gives the validity of the fractal dimension $d_f$.  The value of the slope gives $1-d_f$, also given in Table I. The line fits are, from top to bottom on the left, for the Atat\"urk Dam Lake, Mediterranean, Agean, Black Sea, Bosphorus Europe and Asia, Van Lake, Bosphorus Asia, Bosphorus Europe, Thracian.  Notice how the shores of the Atat\"urk Dam Lake and (to a lesser extent) the Van Lake stand out by their slope and, therefore, by their high fractal dimension $d_f$, which has a geological explanation.}
\end{figure}

The fractal dimension $d_f$ gives the amount of material in an object as function of its linear extent:  If the linear extent is changed by a factor of $b$, the amount of material changes by a factor of $b^{d_f}$.  Thus, physically, the fractal dimension $d_f$ subsums important structural and historical information on the object.  We thus expect the fractal dimension of a shoreline (or any line boundary) to be between 1 (a straigh line) and 2 (a curve compactly covering a surface), and to reflect important information.  We have thus calculated the shores of Turkey:  The shores of the Mediterranean, the Aegean, the Black Sea, the Bosphorus Straits (on both the Asian and European sides), the Van Lake, and the lake formed by the Atat\"urk Dam.  As we shall see below, we have indeed found distinctive results, leading to cogent explanations and associations.

\section{Method}

On a given shore line between two specific points, we would expect
\begin{equation}
L = \lim_{G\rightarrow 0} N(G)\cdot G,
\end{equation}
where $G$ is the length of the unit ruler used in the measurement, $N(G)$ is the number of unit rulers spanning the shore between the two points, and $L$ is the actual shore distance between the two points.  However, in his classic work on the border between Portugal and Spain, Richardson \cite{Richardson, Tarboton} found that $N(G)\cdot G$ did not converge in the limit $G \rightarrow 0$, but that
\begin{equation}
M = \lim_{G\rightarrow 0} N(G)\cdot G^{d_f}
\end{equation}
did converge.  Subsequently, Mandelbrot \cite{Mandelbrot} interpreted $d_f$, generally a non-integer number, as the fractal dimension of the shore line.  Simply set, this is the consequence of the shore line not being comprised, at any length scale, of consecutive small linear units.

Substituting Eq.(1) into Eg.(2),
\begin{equation}
\log L = \log M + (1-d_f) \log G.
\end{equation}
The fractal dimension $d_f$ is found by fitting the slope of this function for varying $G$.

\begin{table}
\begin{tabular}{c c c c c c c c c c c c c c c c}
\hline
\vline &Shores &\vline & $N_1,N_2,N_3$ &\vline &  $d_f$  &\vline &Line Goodness R&\vline \\
\hline \hline
\vline &Mediterranean &\vline & 416,244,180 &\vline &  1.21  &\vline &0.9489 &\vline \\
\hline
\vline &Aegean &\vline & 391,234,170 &\vline &  1.20  &\vline &0.9787 &\vline\\
\hline
\vline &Black Sea &\vline & 224,144,106 &\vline &  1.08  &\vline &0.9963 &\vline\\
\hline
\vline &Bosphorus Asia &\vline & 96,61,43 &\vline &  1.16  &\vline &0.9854 &\vline\\
\hline
\vline &Bosphorus Europe  &\vline & 85,56,41 &\vline &  1.05  &\vline &0.9576 &\vline\\
\hline
\vline &Bosphorus Eur + Asia &\vline & 181,117,84 &\vline &  1.11  &\vline &0.9807 &\vline\\
\hline
\vline &Thrace &\vline & 87,55,41  &\vline & 1.09  &\vline &0.9453 &\vline\\
\hline
\vline &Van Lake &\vline & 136,85,55 &\vline & 1.31  &\vline &0.9492 &\vline\\
\hline
\vline &Atat\"urk Dam Lake &\vline & 485,252,170 &\vline & 1.51  &\vline &0.9914 &\vline\\
\hline \hline
\end{tabular}
\caption{Calculated fractal dimensions $d_f$ of the shore lines of the Mediterranean, the Aegean, the Black Sea, the Bosphorus Straits (on both the Asian and European sides), the Van Lake, and the lake formed by the Atat\"urk Dam, using unit lengths of $G_1,G_2,G_3=1,1.5,2$ cm.  By being close to 1, the fit measure $R$ shows the goodness of the linear fits, also seen in Fig. 2. The number of measurement points on each shore, giving the number of units rulers spanning the shore, are given by $N_1,N2,N3$ for the unit lengths of $1, 1.5,$ and 2 cm, respectively.}
\end{table}

\section{Application: Distinctive Fractal Dimensions}

We have calculated by this method to obtain the fractal dimensions of the different outer and inner shores of Turkey. Maps of different sizes appropriate to the shore object are readily available at www.google.com/maps/@39.1831645,35.0534656,6z. On these maps, we have measured the shore lines of the Mediterranean, the Aegean, the Black Sea, the Bosphorus Straits (on both the Asian and European sides), the Van Lake, and the lake formed by the Atat\"urk Dam, using unit lengths of $G = 1, 1.5,$ and 2 cm.

The Mediterranean was calculated from Dat\c{c}a to the Syrian border, with the number of measurement points on the map, giving the number of units rulers spanning the shore, being $N_1=416, N_2=244, N_3=180$, for the unit lengths of $G = 1, 1.5,$ and 2 cm, respectively.  The (large) numbers of measurement points for each shore are given in Table I.  The Aegean was calculated from Dat\c{c}a to \c{C}anakkale.  The Black Sea was calculated from the Bulgarian border to the Georgian border.  The totality of the Van Lake and Atat\"urk Dam Lake shore lines were calculated.  For the Atat\"urk Dam Lake shore, there were $N_1=485, N_2=252, N_3=170$ measurement points, as explained above.  The larger difference between successive unit lengths is a reflection of the high fractal dimension, as seen below. In addition, the Thracian shore line was calculated the Greek border to Sedd\"ulbahir.  The results are given in Fig. 2 and Table I.

\section{Discussions: Origins and Geology}

The goodness of the linear fits, as seen from Fig. 2 and the last column of Table I, clearly shows the validity of the concept of fractal dimension, which indeed turns out to be more than 1 (a line) and less than 2 (a surface).  Furthermore, our specific results lead to cogent explanations.  The differences in the fractal dimensions clearly reflect the formation history of these boundaries.  The fractal dimension of the Atat\"urk Dam Lake clearly stands out with the maximal value of $d_f=1.5$.  Interestingly, this fractal dimension has the best goodness of linear fit value, $R=0.9914$. This distinctively high value of the Atat\"urk Dam Lake fractal dimension is consistent with the knowledge that this Lake recently formed by flooding meandering and multiply branched rivers.  The fractal dimension of Van Lake also stands with the large value of $d_f=1.3$.  This is supported by the difference in the morphotectonic patterns of these regions as outlined in the following paragraphs.

Anatolia is one of the most strongly deformed continental regions of the World. This is manifested by two geological features: 1-Morphology 2-Earthquakes (Fig.1). Therefore, the landforms are young, formed primarily after the Late Miocene. The two mountain ranges, the Pontide and the Taurus in the North and the South respectively lying along with the shores, rise steeply like a Wall and separate the Central Anatolian Plateau from the sea realm. The coastal regions are tectonically very active and display zigzagging patterns formed as a result of the conjugated pairs of faults of medium (1-10 km) and big (10-100 km) scale \cite{Yilmaz1,Yilmaz2,Yilmaz3,Yilmaz4}.  The Lake Van on the other hand represents an erosional flatland on the elevated Eastern Anatolian High Plateau, which is later filled with water when the broad valley floor was dammed by edifices of the young volcanoes, i.e., the Nemrut and Kirkor volcanoes \cite{Yilmaz1,Yilmaz2,Yilmaz3,Yilmaz4}.

All other fractal dimensions of the shore lines are about $d_f=1.2$.  This consistency in itself is an important fact.  The somewhat lower value of the fractal dimension, $d_f=1.08$, of the Black Sea shore can be understood by the mountain range singularly closely parallel to the shore and in fact making the land mass rather inpenetrable from the narrow coastline. Finally, one would wonder that the Bosphorus was also the result of the flooding of a meandering river \cite{Yilmaz1,Yilmaz2,Yilmaz3,Yilmaz4}, some 8,000 years ago, but does not show the high fractal dimension.  The explanation could be that the Bosphorus is singularly lacking in important branches.  Therefore, the Bosphorus represents an ancient meandering river valley which was flooded by the sea from the Black Sea about 8000 years ago \cite{Yilmaz1,Yilmaz2,Yilmaz3,Yilmaz4}.

\section{Conclusion}

It is seen that fractal dimensions can easily yield important classifications and origin information for shore lines.

\begin{acknowledgments}
We dedicate this paper to the memory of Dietrich Stauffer.  One of us (ANB) was especially fortunate to sit next to him during a course, in his first year as a doctorate student, now 50 years ago. This started a long period of friendship, appreciation, and knowledge exchange.

Support by the TEBIP High Performers Program of the Board of Higher Education of Turkey and by the Academy of Sciences of Turkey (T\"UBA) is gratefully acknowledged.
\end{acknowledgments}


\begin{references}

\bibitem{Richardson} L. F. Richardson, The problem of contiguity: An appendix to statistics of deadly quarrels, General System Yearbook, {\bf6}, 139-187 (1961).
\bibitem{Tarboton} D. G. Tarboton, R. L. Bras, I. Rodriguez-Iturbe, The fractal nature of river networks, Water Resources Research, {\bf24} (8), 1317-1322 (1988).
\bibitem{Mandelbrot} B. Mandelbrot, How long is the coast of Britain? Statistical self-similarity and fractional dimension, Science, New Series {\bf156} (3775), 636-638 (1967).

\bibitem{Yilmaz1} Y. Y{\i}lmaz, Y. G\"uner, F. \c{S}aro\u{g}lu, Geology of the quaternary volcanic centres of the east Anatolia, J. Volcanol. Geotherm. Res. {\bf85}, 173-210 (1998).

\bibitem{Yilmaz2} Y. Y{\i}lmaz, E. G\"oka\c{s}an, A. Y. Erbay, Morphotectonic development of the Marmara region, Tectonophysics {\bf488}, 51–70 (2010).

\bibitem{Yilmaz3} \.I. \c{C}emen and Y. Y{\i}lmaz, eds., Active global seismology-neotectonics and earthquake potential of the Eastern Mediterranean region, Geophysical Monograph {\bf225}, American Geophysical Union, Wiley Press, 306 p., ISBN: 978-1-118-94498 (2017).

\bibitem{Yilmaz4} Y. Y{\i}lmaz, Morphotectonic development of Anatolia and the surrounding regions, in Active global seismology-neotectonics and earthquake potential of the Eastern Mediterranean region, \.I. \c{C}emen and Y. Y{\i}lmaz, eds., Geophysical Monograph {\bf225}, American Geophysical Union, Wiley Press, ISBN: 978-1-118-94498, p.11-91 (2017).












\end{references}
\end{document}